%% file: main.tex
\documentclass[sigconf]{acmart}
\AtBeginDocument{%
  \providecommand\BibTeX{{%
    \normalfont B\kern-0.5em{\scshape i\kern-0.25em b}\kern-0.8em\TeX}}}

\usepackage{amsmath,amsfonts}
\usepackage{algorithmic}
\usepackage{tcolorbox}
\usepackage{graphicx}
\usepackage{listings}
\usepackage{makecell}
\usepackage{adjustbox}
\usepackage{multirow}
\usepackage{graphicx}
\usepackage{svg}
\usepackage{xspace}
\usepackage{dsfont}
\usepackage{lipsum}
\usepackage{subfigure}
\usepackage{centernot}
\usepackage{diagbox}
\usepackage{tabularx}

\newcolumntype{P}[1]{>{\centering\arraybackslash}p{#1}}
\usepackage[ruled,vlined,linesnumbered]{algorithm2e}

\SetCommentSty{mycommfont}
\usepackage{pifont}
\usepackage{enumitem}

\usepackage{caption}
\newcommand{\xmark}{\ding{55}}
\newcommand{\cmark}{\ding{51}}
\newcommand{\tool}{{\texttt{AdbGPT}}\xspace}

\usepackage{xcolor}
\usepackage{xcolor,colortbl}
\definecolor{lightgray}{gray}{0.93}
\definecolor{slightgray}{gray}{0.98}
\definecolor{darkgray}{gray}{0.77}

\definecolor{amber}{rgb}{1.0, 0.49, 0.0}

\newcommand{\chen}[1]{\textcolor{red}{#1}}

\usepackage{framed}
\definecolor{formalshade}{rgb}{0.95, 0.95, 1}
\definecolor{mygray}{gray}{0.4}
\definecolor{lightgray}{gray}{0.93}

\newenvironment{formal}{%
  \MakeFramed{\advance\hsize-1\width\FrameRestore}%
  \vspace{2pt}
  \noindent\hspace{-3pt}
}
{%
  \vspace{2pt}\endMakeFramed%
}

\setcopyright{acmcopyright}
\copyrightyear{2024}
\acmYear{2024}
\setcopyright{acmlicensed}\acmConference[ICSE 2024]{2024 IEEE/ACM 46th International Conference on Software Engineering}{April 14--20, 2024}{Lisbon, Portugal}
\acmBooktitle{2024 IEEE/ACM 46th International Conference on Software Engineering (ICSE 2024), April 14--20, 2024, Lisbon, Portugal}
\acmPrice{15.00}
\acmDOI{10.1145/3597503.3608137}
\acmISBN{979-8-4007-0217-4/24/04}

\begin{document}

\title{Prompting Is All You Need: Automated Android Bug Replay with Large Language Models}

\author{Sidong Feng}
\affiliation{%
  \institution{Monash University}
  \city{Melbourne}
  \country{Australia}}
\email{sidong.feng@monash.edu}

\author{Chunyang Chen}
\affiliation{%
  \institution{Monash University}
  \city{Melbourne}
  \country{Australia}}
\email{chunyang.chen@monash.edu}

\renewcommand{\shortauthors}{Feng et al.}

\begin{abstract}
  Bug reports are vital for software maintenance that allow users to inform developers of the problems encountered while using the software. As such, researchers have committed considerable resources toward automating bug replay to expedite the process of software maintenance. Nonetheless, the success of current automated approaches is largely dictated by the characteristics and quality of bug reports, as they are constrained by the limitations of manually-crafted patterns and pre-defined vocabulary lists. Inspired by the success of Large Language Models (LLMs) in natural language understanding, we propose \tool, a new lightweight approach to automatically reproduce the bugs from bug reports through prompt engineering, without any training and hard-coding effort. \tool leverages few-shot learning and chain-of-thought reasoning to elicit human knowledge and logical reasoning from LLMs to accomplish the bug replay in a manner similar to a developer. Our evaluations demonstrate the effectiveness and efficiency of our \tool to reproduce 81.3\% of bug reports in 253.6 seconds, outperforming the state-of-the-art baselines and ablation studies. We also conduct a small-scale user study to confirm the usefulness of \tool in enhancing developers' bug replay capabilities.
\end{abstract}



\maketitle

\input{introduction}
\input{background}

\input{approach}

\input{eval_rq1}
\input{eval_rq2}

\input{eval_rq3}

\input{userstudy}

\input{threats}
\input{related}

\input{conclusion}

\bibliographystyle{ACM-Reference-Format}
\bibliography{main}

\end{document}

%% file: introduction.tex
\section{Introduction}
\label{sec:introduction}
Mobile applications have gained great popularity in recent years~\cite{chen2019gallery,feng2022gallery,chen2020lost}.
There have been over 3.8 million Android apps and 2 million iPhone apps striving to gain users on Google Play and Apple App Store~\cite{web:appstat}.
Once an app is released, its quality is largely ensured by continuing maintenance activities and one of the most important maintenance tasks is to handle bug reports, i.e., documents that describe software bugs with information about what happened and what the user expected to happen.
Along with the observed and expected behavior, bug reports often go on to contain the steps to reproduce (S2Rs) the bugs that assist developers to replicate and rectify the bugs, albeit with considerable amounts of engineering effort~\cite{planning2002economic}.


In an effort to accelerate bug maintenance, numerous researchers have worked toward providing automated solutions for bug reproduction.
Previous studies~\cite{fazzini2018automatically, zhao2019recdroid, zhao2022recdroid+,liu2020automated} apply natural language processing (NLP) and machine learning (ML) techniques to extract the S2R entities (i.e., action type, target component, input value, and scroll direction) from the bug reports, and employ random or simple guided exploration algorithms for bug reproduction. 
Unfortunately, in many cases, the S2Rs in the bug reports present significant challenges to previous automated approaches of bug replay, and can even be difficult for professional developers to reproduce manually~\cite{aranda2009secret,bettenburg2008extracting,johnson2022empirical,bettenburg2008makes,zhou2012should}.
First, the S2Rs are often unclear, imprecise, and ambiguous, owing to the cognitive and lexical gap between users and developers.

\begin{formal}
    s1. \hspace{0.1cm} Open bookmark \\
    s2. \hspace{0.1cm} Tap ``add new bookmark'' and \colorbox{red!25}{create} a name with ``a'' \\
    s3. \hspace{0.1cm} \colorbox{red!25}{Create} another one with name ``b'' \\
    s4. \hspace{0.1cm} Click ``a'' \\
    s5. \hspace{0.1cm} Go back to bookmark \colorbox{blue!25}{after} changing name ``a'' to ``b'' \\
    s6. \hspace{0.1cm} App crash
\end{formal}

Given the S2R shown above, extracting entities from the S2R requires a robust semantic and syntactic understanding.
For instance, the two ``create'' words in \textit{s2} and \textit{s3} are not semantically identical; one refers to ``input'' while the other refers to ``repeat'' the previous steps.
In \textit{s3}, to ``create another one'', a deep comprehension of the specific creation steps from the previous steps is needed.
Furthermore, a step may contain multiple sub-steps, such as \textit{s5}, which can be divided into ``go back to bookmark'' and ``change name `a' to `b'\,''.
Last but not least, in \textit{s5}, the conjunction word ``after'' alters the temporal order of the steps, meaning ``change name `a' to `b'\,'' should be executed first, followed by ``go back to bookmark''.
These challenges surpass pre-defined word lists and are difficult to address using specific patterns based on previous works.
Second, the S2Rs are often incomplete, and developers from more than 1.3k open-source projects wrote a letter to GitHub expressing their frustration of the S2Rs are often missing in the bug reports~\cite{web:letter}, and requested a solution that would encourage users to include them.
So much so that the previous automated approaches are unable to replicate the bugs. 

Emerging Large Language Models (LLMs), such as GPT-3~\cite{brown2020language}, PaLM~\cite{chowdhery2022palm}, RoBERTa~\cite{liu2019roberta}, T5~\cite{raffel2020exploring}, have been trained on ultra-large-scale corpora and exhibit promising performance in natural language understanding and logical reasoning.
For example, GPT-3~\cite{brown2020language} (Generative Pre-trained Transformer3) from OpenAI with 175 billion parameters trained on billions of resources can smartly retrieve media information~\cite{taecharungroj2023can,zhang2023prompting}, guide cloze selection~\cite{onoe2022entity}, etc.
The recent success of ChatGPT~\cite{web:chatgpt} based on GPT-3.5 demonstrates its remarkable ability to comprehend human questions and act as a knowledgeable assistant for interacting with users.
Inspired by the impressive advancements of LLMs in the fields of information retrieval, selection guidance, and role-playing, we designate the LLMs as expert developers capable of extracting entities from bug reports and guiding replays from a set of dynamic GUI components.

In this paper, we propose a novel approach called \tool to \textbf{A}utomatically repro\textbf{D}uce \textbf{B}ugs using LLMs.
\tool consists of two phases: i) \textit{S2R Entity Extraction} and ii) \textit{Guided Replay}.
Notably, the underlying approach of \tool for these two phases is prompt engineering, i.e., prompting the tasks to generate desired output, which is extremely lightweight compared to NLP with manually-crafted patterns and ML with massive training data.
To extract the S2R entities, we first provide the LLMs with the knowledge of entity specifications, including available actions and action primitives.
Since the LLMs are not specifically designed to handle bug reports, we employ few-shot learning by giving a few representative examples to help LLMs recognize the task of S2R entity extraction.
Additionally, we provide LLMs with detailed chain-of-thought reasoning from developers, endowing it to extract S2R entities with the same thought process as developers.
Given the inferred S2R entity extraction from the LLMs, we next prompt LLMs to dynamically guide the replay according to the GUI screens.
To provide the current GUI information to the LLMs, we propose a novel GUI encoding algorithm to ensure the integrity of the information and the effectiveness of encoding.
By giving a few examples with intermediate reasoning, LLMs infer the target components to operate on the GUI screen, repeating the steps to trigger the bugs.

We conduct a comprehensive evaluation to measure the effectiveness and efficiency of our \tool.
First, we evaluate the performance of our \tool in extracting S2R entities from 88 bug reports.
Compared with two state-of-the-art baselines and two ablation studies, our approach achieves significantly higher accuracy, i.e., 90.4\% and 90.8\% in step extraction and entity extraction, respectively.
Second, we evaluate the performance of our \tool in guiding bug replay and show that our approach can successfully reproduce 81.3\% of bugs, outperforming the baselines and ablations.
In addition, our approach is much more efficient compared with the baselines, saving an average of 1105.17 seconds per bug report.
Apart from the performance of our tool, we also assess the perceived usefulness of our \tool by conducting a small-scale user study on replaying bugs from 5 real-world bug reports.
Through the study, we provide initial evidence of the usefulness of \tool in facilitating bug replay.
The contributions of this paper are as follows:
\begin{itemize}
    \item This is the first work to exploit LLMs into bug report analysis and GUI guidance, paving the way for new opportunities in software engineering tasks. 
    \item We propose a lightweight approach, \tool\footnote{https://github.com/sidongfeng/AdbGPT}, that utilizes prompt engineering with few-shot learning and chain-of-thought reasoning to harness LLMs' knowledge for automated bug replay.
    \item Comprehensive experiments, including the performance evaluation of \tool and its detailed qualitative analysis, reveal the capabilities of LLMs in bug replay. A user study to further demonstrate the usefulness of our approach.
\end{itemize}

%% file: background.tex
\section{Background}
We briefly discuss Large Language Models (LLMs), and in-context learning and chain-of-thought reasoning that we adapt in \tool.

\begin{figure}
	\centering
	\includegraphics[width = \linewidth]{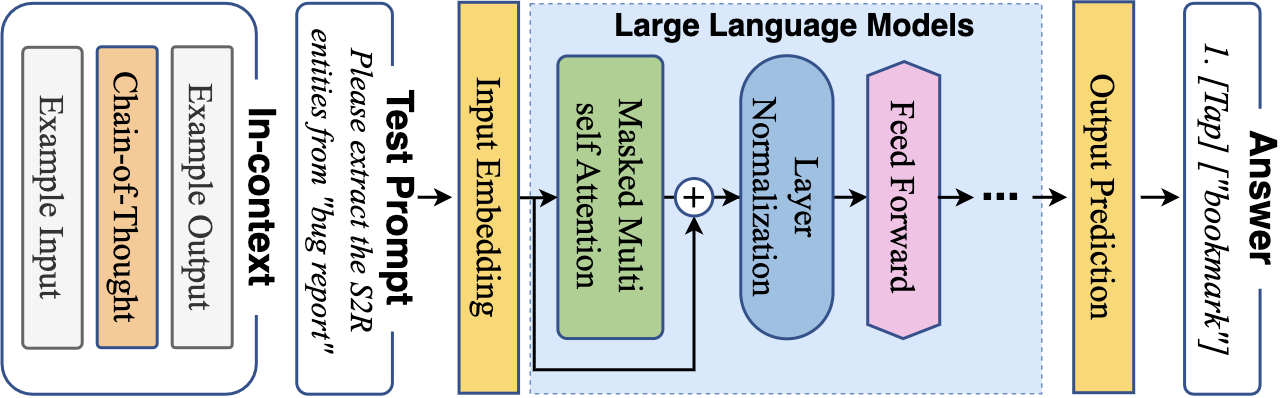}
	\caption{The process of prompt engineering.}
	\label{fig:background}
\end{figure}

\subsection{Large Language Models}
Pre-trained Large Language Models (LLMs) introduce a new era in natural language understanding.
It is trained on ultra-large-scale corpora and can support a wide range of natural language processing tasks with prompt engineering.
By simply describing natural language prompts, LLMs can be invoked to perform specific tasks without requiring additional training or hard coding.
Figure~\ref{fig:background} illustrates the process of prompt engineering.
Given an input/test prompt (sentence with task-specific instructions), the input embedding layer encodes it through word embedding. 
To comprehend prompts and generate corresponding answers, LLMs employ a Transformer model~\cite{vaswani2017attention}.
In detail, the multi-self attention layer is used to divide a whole high-dimensional space into several different subspaces to calculate the similarity. 
The normalization layer is implemented through a normalization step that fixes the mean and variance of each layer’s inputs. 
The feed-forward layer compiles the data extracted by previous layers to form the final answer.
In this work, we prompt the LLMs to extract S2R entities and generate dynamic actionable operations, and decode the LLMs' feedback to reproduce bugs.

\subsection{In-context Learning}
One common paradigm to master a specific task is to fine-tune the models~\cite{lecun2015deep}.
The fundamental of model fine-tuning is big data, which is labor-intensive and costly.
With the increasing ability of LLMs, in-context learning has shifted to a new paradigm, known as zero-shot learning, where LLMs make predictions by directly describing the desired output.
While zero-shot learning shows promising performances in various training tasks by leveraging prior knowledge from training resources, it remains challenging to apply to unseen tasks~\cite{fan2022automated,brown2020language,jiang2019inferring}.
To overcome this challenge, few-shot learning is utilized to augment the context with a few examples of desired inputs and outputs (see Figure~\ref{fig:background}).
This enables LLMs to recognize the input prompt syntax and patterns of the output.
In our work, as LLMs are not specifically trained to understand, analyze, organize, and reproduce bug reports, we adopt few-shot learning as the in-context paradigm to help LLMs extract S2R entities and guide bug replay.

\subsection{Chain-of-Thought Reasoning}

While few-shot learning has proven effective for simple tasks with a few examples of \textit{<input, output>}, it faces difficulties with more complicated tasks that require logical thinking and multiple steps to solve, such as arithmetic or commonsense reasoning questions~\cite{kojima2022large,wang2022rationale, zhou2022least,ling2017program}.
To elicit the capability of LLMs' reasoning, many researchers pioneer the idea of using rationales as the intermediate steps in examples.
This idea is formulated as chain-of-thought reasoning~\cite{wei2022chain}, i.e., \textit{<input, reasons, output>} in Figure~\ref{fig:background}.
Our task of extracting S2R entities and guiding bug replay is not simple, it requires logical thinking to understand the bug reports and a coherent series of intermediate steps to trigger the bugs.
As a result, we employ chain-of-thought reasoning to endow the LLMs to reproduce the bugs through the minds of expert developers.

%% file: approach.tex
\section{\tool Approach}
\begin{figure}
	\centering
	\includegraphics[width = 0.8\linewidth]{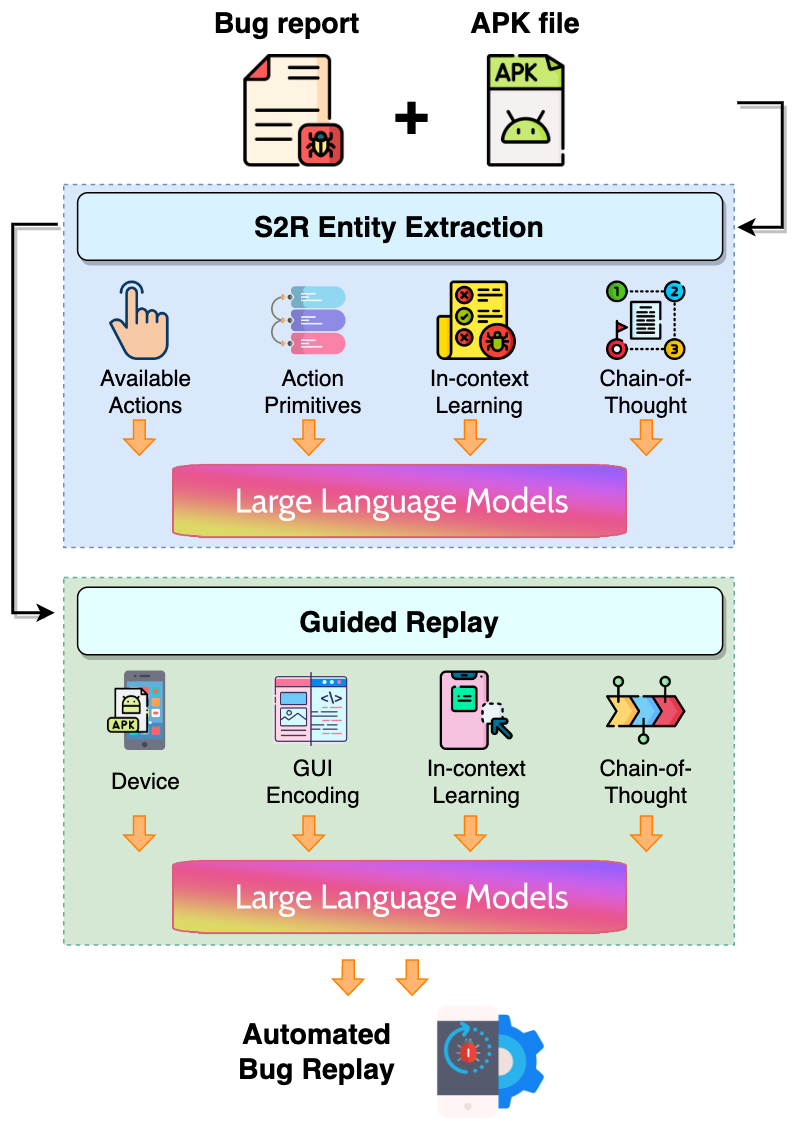}
	\caption{The overview of \tool.}
	\label{fig:overview}
\end{figure}

Given a bug report and the app, we propose an automated approach to extract the S2R entities and reproduce each step based on the current GUI state to trigger the bugs in the app.
The overview of our approach is shown in Figure~\ref{fig:overview}, which is divided into two main phases: 
(i) the \textit{S2R Entity Extraction} phase, which extracts the S2R entities defining each step to reproduce the bug report, including action types, target components, input values, or scroll directions;
(ii) the \textit{Guided Replay} phase that matches the entities in S2R with the GUI states to repeat the bug reproduction steps.
While many works~\cite{zhao2019recdroid,zhao2022recdroid+,fazzini2018automatically,liu2020automated} attempt to address these two phases using different ad-hoc methods, such as natural language processing for extracting S2R entities and greedy algorithms for guiding replay, our approach is notably lightweight compared to them, we exploit the single LLMs to address both phases through novel prompt engineering.

\subsection{S2R Entity Extraction}
\label{sec:phase1}
The first phase of our approach is to understand, analyze, organize, and extract the entities from the S2R text in a bug report.
Specifically, we leverage LLMs, equipped with knowledge learned from large-scale training corpora, to provide an in-depth understanding of the potential entities presented in the bug report.
As discussed in Section~\ref{sec:introduction}, a step can be expressed in various ways and with different words.
Therefore, we first provide LLMs with information about entity specifications, including available actions and action primitives, which can be mapped to instructions for device replay.
To help LLMs recognize our task and infer S2R entities, we provide a few examples of conditioning, such as an S2R as input, a chain-of-thought as reasoning, and the final entities as the output.
An example of the prompts is shown in Table~\ref{tab:prompt1}.
With this in-context few-shot learning, we prompt the LLMs in the same way to extract the S2R entities from the test bug report.

\subsubsection{Available actions}
\label{sec:actions}
Unlike target components and input values, which are indeterministic and dynamic depending on the current GUI, actions that can interact with the device remain consistent.
We identify five standard actions as shown in Table~\ref{tab:prompt1}, including \textit{Tap, Scroll, Input, Double-tap}, and \textit{Long-tap}.
While there are other customized actions, such as pinch, multi-handed gestures, etc., they are less prevalent.
For brevity, we focus on the commonly-used actions in this paper.

\renewcommand{\arraystretch}{1.15}
\begin{table}
        \footnotesize
	\centering
	\caption{The example of prompt engineering of S2R entity extraction. }
	\label{tab:prompt1}
	\begin{tabularx}{\linewidth}{l|X} 
		\hline
		\rowcolor{darkgray} \bf{PROMPT TYPE}  & \bf{INSTANTIATION} \\
		\hline
		\hline
		 Available actions & \textit{Tap, Scroll, Input, Double-tap, Long-tap} \\
		  \hline
		   Action primitives & \textit{[Tap] [Component], \hspace{0.03cm} [Scroll] [Direction], \hspace{0.03cm} [Input] [Component] [Value], \hspace{0.03cm} [Double-tap] [Component], \hspace{0.03cm} [Long-tap] [Component]} \\
		  \hline
		   Example input & \textit{\makecell[Xt]{1. Open bookmark \\ 2. Tap ``add new bookmark'' and create a name with ``a'' \\ 3. Create another one with name ``b'' \\ 4. Click ``a'' \\ 5. Go back to bookmark after changing name ``a'' to ``b'' \\ 6. App crash}} \\
		\hline
		  Chain-of-Thought & \textit{\makecell[Xt]{\textbf{1st} step is ``Open bookmark''. The action is ``open'' and the target component is ``bookmark''. However, there is no explicit ``open'' in the Available actions list. Therefore, we select the closest semantic action ``tap''. Following the Action primitives, the entity of the first step is [Tap] [``bookmark'']. \\ 
          \vspace{0.15cm}
          \textbf{2nd} step is ``Tap `add new bookmark' and create a name with `a'\,''. Due to the conjunction word ``and'', this step can be separated into two sub-steps, ``Tap `add new bookmark'\,'' and ``create a name with `a'\,'' ... \\ 
          \vspace{0.15cm}
          \textbf{3rd} step is ``Create another one with name `b'\,''. Due to its semantic meaning, this step is meant to repeat the previous steps to add another bookmark with name ``b''. Therefore, it should actually be the 2nd step ... \\
          \vspace{0.15cm}
          \textbf{4th} step is ``Click `a'\,'' ... \\
          \vspace{0.15cm}
          \textbf{5th} step is ``Go back to bookmark after changing name `a' to `b'\,''. Due to the conjunction word ``after'', this step can be separated into two sub-steps, ``Go back to bookmark'' and ``change name `a' to `b'\,''. The conjunction word ``after'' also alters the temporal order of the sub-steps, that ``change name `a' to `b'\,'' should be executed first, followed by ``go back to bookmark'' ... \\
          \vspace{0.15cm}
          \textbf{6th} step is ``App crash''. This step does not have any operations.}} \\
		\hline
             Example output & \textit{\makecell[Xt]{Overall, the extracted S2R entities are: \\ 1. [Tap] [``bookmark''] \\ 2. [Tap] [``add new bookmark''] \\ 3. [Input] [``name''] [``a''] \\ 4. [Tap] [``add new bookmark''] \\ 5. [Input] [``name''] [``b''] \\ 6. [Tap] [``a''] \\ 7. [Input] [``name''] [``b''] \\ 8. [Tap] [``back'']
            }}  \\
            \hline
	\end{tabularx}
\end{table}

\subsubsection{Action primitives}
As the context of action varies, each action requires a different set of primitives to represent entities.
For example, the \textit{Tap} action requires a target component to interact with, such as a button in the GUI.
As a result, we formulate it as \textit{[Tap] [Component]}.
Similarly, for \textit{Double-tap} and \textit{Long-tap} actions, we formulate them using similar linguistic primitives.
The \textit{Scroll} action requires indicating the direction of the scrolling effect, such as upward or downward, formulating as \textit{[Scroll] [Direction]}.
The \textit{input} action involves the process of entering a specific value into a text field component.
To formulate this action, we use the primitive \textit{[Input] [Component] [Value]}.

\subsubsection{In-context learning}
\label{sec:phase1_incontext}
A representative example helps the model elicit specific knowledge and abstractions needed to complete the task.
A common strategy for example selection is to randomly sample from a dataset~\cite{wang2020generalizing}.
However, a random bug report may not encompass all the complexities of the task, thereby limiting the ability of LLMs to gain a comprehensive understanding of the problem.
To select representative examples, we recruit three professional developers with four-year experience in bug fixing and triaging by word-of-mouth.
First, we give them an introduction to our study and ask them to interact with the LLMs to become familiar with our task.
Each developer is then asked to independently review a large-scale of bug reports collected from previous works~\cite{feng2022gifdroid,feng2022gifdroid1} and select the challenging ones.
After 1.5 hours of initial reviewing, each developer on average examines 110 bug reports and selects 7 as challenging.
The developers then meet and discuss the representatives of the selection and refine the dataset until a consensus is reached.
In total, we collect 15 bug reports as our representative dataset, with an example in Table~\ref{tab:prompt1}.

\subsubsection{Chain-of-thought reasoning}
\label{sec:phase1_cot}
Giving a chain-of-thought reasoning, it endows the LLMs to generate a coherent series of intermediate steps that lead to a reasonable output.
To employ this paradigm, we ask the developers to independently write the chain-of-thought reasoning for each bug report in the representative dataset, following the previous study~\cite{wei2022chain}.
The three developers and the first author then meet and discuss the optimal reasoning for each bug report, aiming to express a clear step-by-step thinking process that leads to the final S2R entities.
An example of chain-of-thought reasoning is shown in Table~\ref{tab:prompt1}.
In the 1st step, we explicitly explain the action (``Open'') and target component (``bookmark'').
Since ``Open'' is not listed in the \textit{Available actions}, we map it to the closest semantic action ``Tap''.
For the 2nd and 5th steps, we explicitly explain the conjunction words that lead to multiple sub-steps and alternate temporal order.
The 3rd step is more complicated, that we explicitly explain the semantic meaning of the step and its relationship to the previous steps.
Overall, this step-by-step thinking leads to the final output of the extracted S2R entities.

\subsubsection{Prompt construction}
We combine the aforementioned information in Table~\ref{tab:prompt1} as the input prompt, i.e., ($\langle$\textit{Available actions}$\rangle$ + $\langle$\textit{Action primitives}$\rangle$ + $\langle$\textit{Example input}$\rangle$ + $\langle$\textit{Chain-of-Thought}$\rangle$ + $\langle$\textit{Example output}$\rangle$). 
Note that, due to the robustness of the LLMs, the prompt sentence does not need to adhere strictly to grammar rules~\cite{brown2020language}.
Next, we input the test bug report as the test prompt and query for the S2R entities.
Due to the advantage of few-shot learning and chain-of-thought reasoning, the LLMs will consistently generate a numeric list to represent the extracted S2R in the same format as our example output, which can be inferred using regular expressions.

\subsection{Guided Replay}
\label{sec:phase2}
The second phase is to explore the apps to match the extracted S2R entities to a sequence of GUI events to reproduce the bugs.
A seemingly straightforward solution~\cite{fazzini2018automatically,zhao2019recdroid,zhao2022recdroid+} is to use lexical computation to match the extracted components against the displayed text of the components on the current GUI screen.
However, as explained in Section~\ref{sec:introduction}, this process of component mapping is limited due to the issues of missing steps, i.e., there are no matching components on the GUI screen.
To address this, we adopt LLMs with few-shot learning and chain-of-thought reasoning to generate dynamic guidance on the GUI screen, enabling automatic reproduction of the steps even when a step is missing.
One challenge in using LLMs for GUI guidance is that they can only process reasonably sized text input.
To help LLMs inherently understand GUI screens, we propose a novel method to transfer the GUIs into domain-specific prompts that the LLMs can understand.
Given each step and its current GUI screen, we prompt the LLMs to generate the target component to operate on, ultimately triggering the bug.

\begin{figure}
	\centering
	\includegraphics[width = 0.995\linewidth]{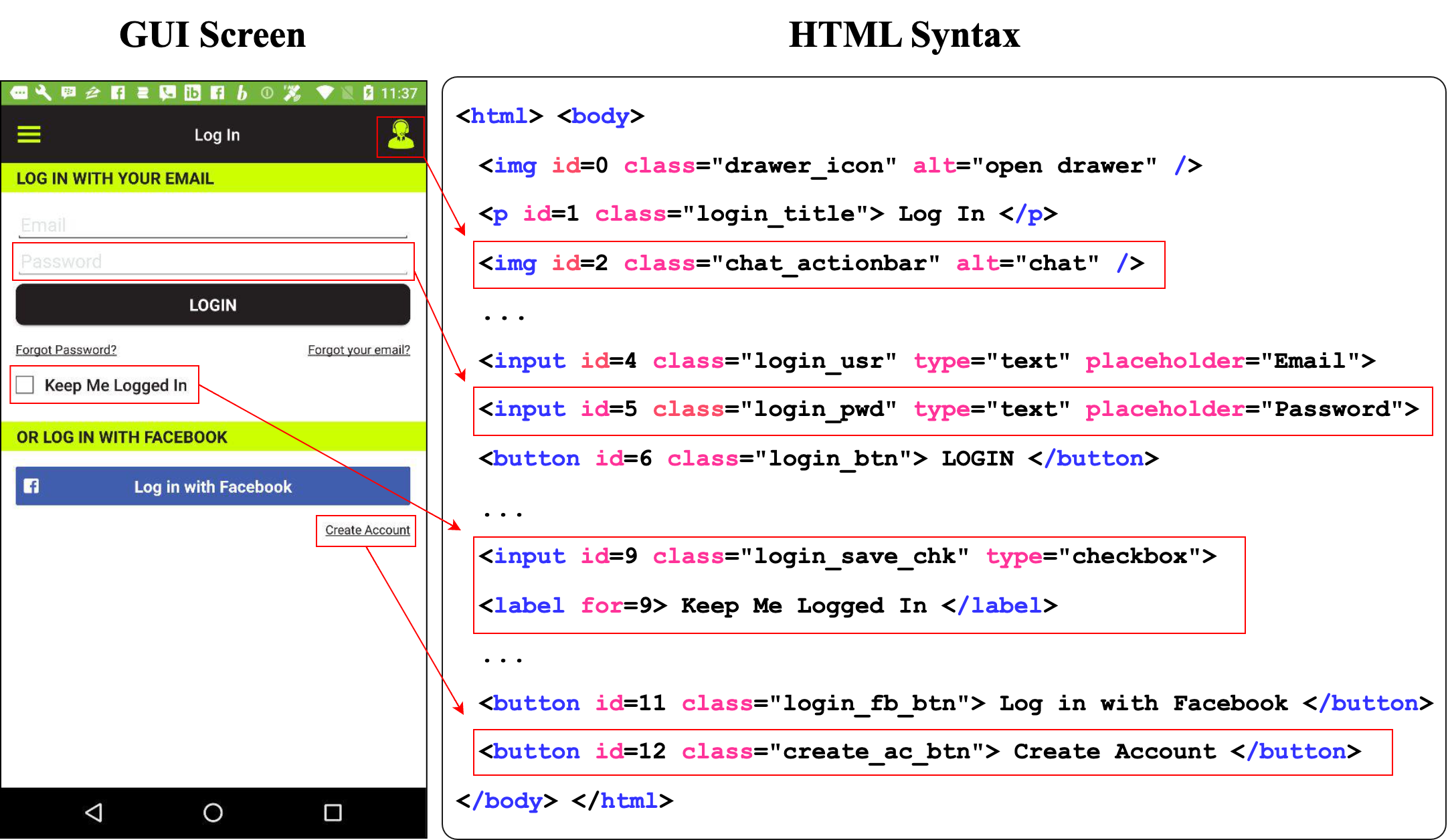}
	\caption{Illustration of GUI encoding.}
	\label{fig:ui}
\end{figure}

\subsubsection{GUI encoding}
To prompt the LLMs to guide bug replay according to the GUIs, we need to dynamically depict the current GUI screen, as well as its contained component information, such as text, type, image, etc.
To design our screen encoding, we leverage the insight that if our GUI encoding falls within the training data distribution of LLMs, it is more likely to perform better~\cite{chen2022knowprompt,liao2022ptau,zhou2022learning}.
Since most of the LLMs' training data is typically scraped from the raw web page~\cite{brown2020language}, we encode a GUI in a text by converting its view hierarchy data into HTML syntax. 
However, there are two fundamental limitations.
First, the native classes in the view hierarchy don't always match HTML tags, for instance, a \textit{<CheckBox>} in the hierarchy corresponds to a combination of \textit{<input type="checkbox">} and \textit{<label>}.
Second, including all properties of GUI components will result in excessively long HTML text that may influence the understanding of LLMs.

Therefore, we adopt a heuristic approach to convert the view hierarchy of the GUI screen into HTML text.
Note that since the view hierarchy is not designed to be encoded in HTML syntax, a perfect one-to-one conversion does not exist. 
Our goal is to make the converted view hierarchy look and function similar to the HTML syntax.
An example of the GUI screen and the encoded HTML text is illustrated in Figure~\ref{fig:ui}.
The encoding is conducted by traversing the view hierarchy tree using a depth-first search. 
We convert each node into HTML syntax, preserving a related subset of properties from the view hierarchy.

\begin{itemize}
    \item \texttt{resource\_id}: the unique resource id of component
    \item \texttt{class}: the native component type
    \item \texttt{text}: the text of component
    \item \texttt{content\_desc}: the description of visual component
\end{itemize}

We first map the classes to HTML tags with similar functionalities using an ad-hoc heuristic method.
For example, the \textit{TextView} is mapped to the \textit{<p>} tag as they both present texts.
Similarly, the \textit{Button} is mapped to the \textit{<button>} tag; and \textit{ImageView} to the \textit{<img>} tag.
For special classes in the view hierarchy (e.g., \textit{<EditText>}, \textit{<CheckBox>}, \textit{<RadioButton>}), we use a combination of \textit{<input>} and \textit{<label>}, aligning with HTML syntax.
Note that we focus on the most commonly-used classes for simplicity, and the rest of the classes, such as \textit{<VideoView>}, are mapped to the \textit{<div>} tag.

Then, we insert the text properties of the component in between the opening and
closing HTML tags, following the standard syntax of texts in HTML.
The resource\_id property usually contains additional descriptions of a component’s functionality or purpose, written by developers.
For example, a resource\_id of ``submit\_btn'' refers to the functionality of submit.
We insert it in the ``class'' attributes in HTML to help the models understand the screen context.
For the \textit{ImageView} component, we further insert the content\_desc as the ``alt'' attribute in the HTML to express the accessibility to the image in the GUI screen.
Lastly, we follow the traversing order in the view hierarchy tree to insert unique numeric indexes to each element as the "id" attribute.

\renewcommand{\arraystretch}{1.15}
\begin{table}
        \footnotesize
	\centering
	\caption{The examples of prompt engineering of guide replay.}
	\label{tab:prompt2}
	\begin{tabularx}{\linewidth}{l|X} 
		\hline
		\rowcolor{darkgray} \bf{PROMPT TYPE}  & \bf{INSTANTIATION} \\
		\hline
            \addlinespace[0.1cm] 
            \multicolumn{2}{l}{\textbf{\small{Example 1}}} \\
            \hline
		GUI encoding & \textit{\makecell[Xt]{<html>\hspace{0.12cm} ... \hspace{0.12cm} </html>}} \\
		  \hline
		Example input & \textit{If I need to [Tap] [``Sign in''], which component id should I operate on the GUI?} \\
		\hline
		  Chain-of-Thought & \textit{\makecell[Xt]{ 
          There is no explicit ``Sign in'' component in the current GUI screen. However, there is a semantic closest component ``Log in'' button. The id attribute of ``Log in'' component is 6. So, we could potentially operate on [id=6] in the screen.}} \\
		\hline
            Example output & \textit{[id=6]}  \\
            \hline

            \addlinespace[0.15cm] 
            \multicolumn{2}{l}{\textbf{\small{Example 2}}} \\
            \hline
		GUI encoding & \textit{\makecell[Xt]{<html>\hspace{0.12cm} ... \hspace{0.12cm} </html>}} \\
		  \hline
		Example input & \textit{If I need to [Tap] [``darkmode''], which component id should I operate on the GUI?} \\
		\hline
		  Chain-of-Thought & \textit{\makecell[Xt]{ 
          There is no explicit and semantic similar ``darkmode'' component in the current GUI screen, so it appears a [MISSING] step. However, ``darkmode'' could be related to the ``display'' button in the screen. The id attribute of ``display'' component is 1. So, we could potentially operate on [id=1] component in the screen.}} \\
		\hline
            Example output & \textit{[MISSING] [id=1]}  \\
            \hline
	\end{tabularx}
\end{table}

\subsubsection{In-context learning}
It can be extremely laborious and time-consuming to manually collect representative examples for guided replay, such as checking whether there is a step missing and identifying the target components that may trigger the bugs.
To this end, we aim to synthesize a few representative examples to help the LLMs understand the challenges of missing steps in the S2Rs. 
To synthesize the missing step, we randomly omit 0 to 2 steps in the bug reports.
Given these examples where steps may or may not be missing, we ask the three developers to validate the examples and select the challenging ones following similar procedures in Section~\ref{sec:phase1_incontext}, resulting in 5 representative examples as the dataset.

\subsubsection{Chain-of-thought reasoning}
We further ask the three developers to write their chain-of-thought reasoning, leading to the final output. 
The documenting process is similar to the previous phase in Section~\ref{sec:phase1_cot}, with independent writing followed by a discussion to select the most representative step-by-step reasoning.
We show two particular reasoning in Table~\ref{tab:prompt2}. 
In Example 1, we explicitly explain a semantic example of a guiding step that does not exactly component match S2R (``Sign in'') and GUI (``Log in'').
In Example 2, we explicitly explain an example of guiding a missing step with a potential target component and a missing tag.

\subsubsection{Prompt construction}
We provide a few examples with their chain-of-thought as input prompt, i.e., ($\langle$\textit{GUI encoding}$\rangle$ + $\langle$\textit{Example input}$\rangle$ + $\langle$\textit{Chain-of-Thought}$\rangle$ + $\langle$\textit{Example output}$\rangle$).
Given each step generated by the S2R entity extraction phase and the current GUI encoding as the test prompt, we then query for the target component selection to reproduce the step.
Note that we infer the target component using ``id'', which is more efficient and space-saving than spelling out the complete HTML tag.
In the scenario of missing steps, the output prompt will additionally predict a ``MISSING'' tag to allow us to iteratively explore the steps in the app until the target component is found.
If LLMs fail to explore a potential target component to recover the missing steps (e.g., no ``id'' in the inference), we go back to the previous step and prompt the models to explore a new component.

\subsection{Implementation}
For the LLMs, we use the state-of-the-art ChatGPT model~\cite{web:chatgpt}.
According to a small-scale pilot study, we set the number of few-shot learning examples in the range of 1 to 3.
This is how many examples can fit in the LLM's implicit maximum input tokens (i.e., 4,096 for ChatGPT).
As the LLMs may generate verbose output (such as repeat questions, chain-of-though reasons, etc.), we use ``[]'' to infer the specific prediction, such as entity for S2R entity extraction, and target component id attribute or missing step flag for guided replay.
Note that, users may describe input actions vaguely, for example, ``enter the name''.
To address this, we set a ``test'' value if there is no explicit input value in the step.

Our \tool is implemented as a fully automated bug replay tool.
In detail, we use Genymotion~\cite{web:genymotion} for running and controlling the virtual Android device, Android UIAutomator~\cite{web:uiautomator} for dumping the GUI view hierarchy, and Android Debug Bridge (ADB)~\cite{web:adb} for replaying the steps.

%% file: eval_rq1.tex
\section{Evaluation}
\label{sec:evaluation}
In this section, we describe the procedure we used to evaluate \tool in terms of its performance automatically.
Since our approach consists of two main phases, we evaluate each phase of \tool, including S2R Entity Extraction (Section~\ref{sec:phase1}) and Guided Replay (Section~\ref{sec:phase2}).

\begin{itemize}[leftmargin=0.3cm] 
    \item \textbf{RQ1:} How accurate is our approach in extracting S2R entities?
    \item \textbf{RQ2:} How accurate is our approach in guiding bug replay?
    \item \textbf{RQ3:} How efficient is our approach in bug replay?
    \item \textbf{RQ4:} How usefulness is our approach for developers in real-world bug replay?
\end{itemize}

For \textbf{RQ1}, we present the general performance of our approach for S2R entity extraction and the comparison with state-of-the-art baselines. 
Besides, we also present the performance comparison among the variations of in-context learning (e.g., few-shot vs zero-shot) and the contribution of reasoning by comparing the performance with and without chain-of-thought.
For \textbf{RQ2}, we carry out experiments to check if our approach can trigger the target components, comparing with the baselines and ablations.
For \textbf{RQ3}, we evaluate the runtime overhead of our approach in bug replay.
For \textbf{RQ4}, we conduct a small-scale user study to evaluate the perceived usefulness of \tool for automatically replaying bug reports in real-world development environments.

\subsection{RQ1: Performance of S2R Entity Extraction}
\label{sec:rq1}
\textbf{Experimental Setup.}
To answer RQ1, we first evaluated the ability of our approach (in Section~\ref{sec:phase1}) to accurately identify the S2R entities from the bug reports.
To avoid potential bias, we collected the bug reports from the artifacts of three existing open-source datasets:
(i) the evaluation dataset of ReCDroid~\cite{zhao2019recdroid, zhao2022recdroid+};
(ii) an empirical study on Android bug report reproduction ANDROR2+~\cite{johnson2022empirical};
and (iii) an empirical study of crash bug reports Themis~\cite{su2021benchmarking}.
Due to some overlap across these datasets, we first removed duplicates if the bug reports were from the same issue repository.
The users may report the S2Rs using visual information, such as screenshots or videos~\cite{feng2022gifdroid}.
In this work, we focused on textual information, i.e., the natural language descriptions of S2Rs, so we further audited the bug reports and retained those only containing textual S2Rs.
In total, we collected 88 bug reports as our experimental dataset.

Given the set of experimental bug reports, we needed to manually identify and generate the ground-truth for S2R entities (specific action types, target components, input values, and scroll direction) from the bug reports, as the state-of-the-art automated approaches still cannot 100\% accurately infer the entities.
We recruited two paid students through the university’s internal slack channel and they were compensated with \$12 USD per hour.
They had experience in Android development and bug reproduction.
To ensure accurate annotations, the process began with initial training. 
First, we provided them with an introduction to our study and an example set of annotated bug reports where the authors had annotated the S2R entities. 
Then, we asked them to pass an assessment test.
The two annotators were then assigned the experimental set of bug reports to label the S2R entities independently, without any discussion. After the initial labeling, the annotators met and sanity corrected any subtle discrepancies. Any disagreements were handed over to the first author for the final decision.
In total, we obtained 305 reproduction steps from 88 bug reports, which contained 305 actions, 291 components, 39 inputs, and 14 directions.

\textbf{Metrics.}
An extracted S2R was determined to be correct if all of the following conditions matched the ground-truth, including steps (i.e., step ordering, sub-step separation) and entities (i.e., the action types of each step, the possible target components if existed, the possible input values, and the possible scroll directions). 
Therefore, we employed the accuracy of the S2R extraction as our evaluation metric.
The higher the accuracy score, the better the approach can identify the steps in the bug reports.

\textbf{Baselines.}
We set up two state-of-the-art methods which are widely used for S2R entity extraction as the baselines to compare with our approach.
\textit{ReCDroid}~\cite{zhao2019recdroid} analyzes the dependencies among words and phrases from hundreds of bug reports and defines 22 grammar patterns to extract the S2R entities.
For example, a noun phrase (NP) followed by a ``click'' action should be the target component, etc.
We adopted their released repository for evaluation.
ReCDroid+~\cite{zhao2022recdroid+} is a later work that scraped bug reports from the issue track websites, which is out of the scope of this study. 
Since these two approaches perform the same for S2R entity extraction, we used the ReCDroid in the rest of the paper for simplicity. 
\textit{MaCa}~\cite{liu2020automated} proposes an automated S2R identification tool, which first identifies action words based on natural language processing and machine learning classifier, then extracts its related entities (i.e., target components).
Note that we did not adopt Yakusu~\cite{fazzini2018automatically} as a baseline, as it was developed five years ago and many of its libraries are no longer maintained.

Apart from the baselines, we also added two derivatives of our approach to demonstrate the impact of prompt engineering. 
In Section~\ref{sec:phase1_incontext}, we utilized few-shot learning to help the LLMs understand the requirement of our task. 
Therefore, we set up an ablation study to infer the results without any examples, named \textit{\tool w/o Few}, which is the well-known method called zero-shot learning. 
To demonstrate the strength of the chain-of-thought outlined in Section~\ref{sec:phase1_cot}, we also conducted an ablation study, namely \textit{\tool w/o CoT}.
Specifically, it provides a few examples of bug reports and the outputs, without any intermediate reasons, and then prompts the models to extract the S2R entities from the test bug report.

\textbf{Results.}
Table~\ref{tab:rq1_performance} depicts the performance of our approach in extracting the S2R entities from the bug reports.
The performance of our approach is significantly better than that of other baselines in all metrics, i.e., on average 39.3\%, and 42.2\% more accurate in step extraction and entity extraction compared with the best baseline (MaCa).
In addition, applying few-shot learning and chain-of-thought which provide examples with intermediate reasons, can endow the LLMs with a better understanding of the task, resulting in a boost of performance from 84.6\%, 87.6\% to 90.8\%.

\begin{figure}
	\centering
	\includegraphics[width = 0.99\linewidth]{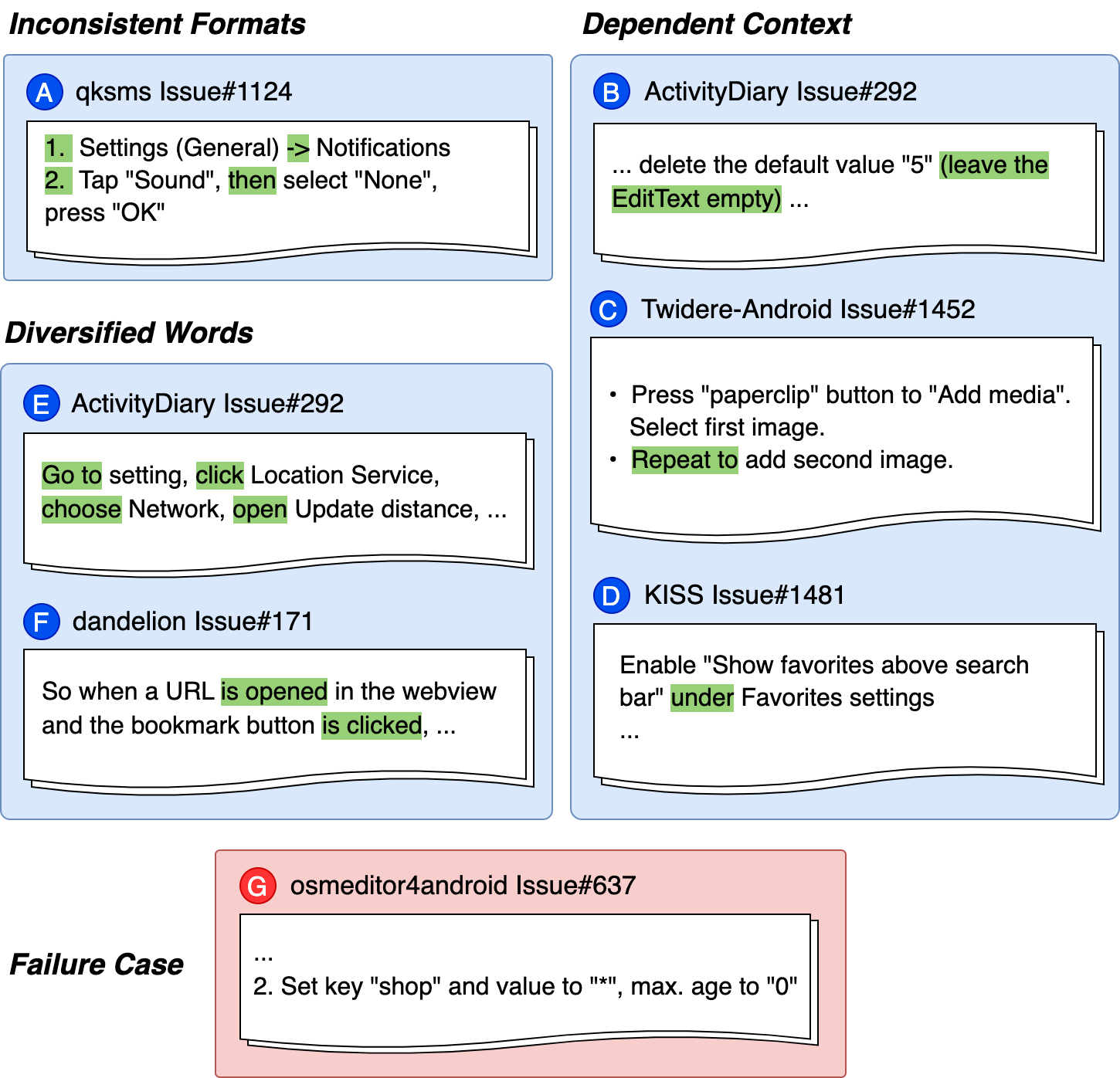}
	\caption{Examples of S2R extraction.}
	\label{fig:rq1}
\end{figure}

To fully explore the reason why our approach outperforms other baselines, we carry out a qualitative study by investigating the bug reports which are correctly extracted by our approach but wrongly extracted by the baselines.
We summarize three reasons, including inconsistent formats, dependent context, and diversified words.
Some representative examples can be seen in Figure~\ref{fig:rq1}.

\textit{1) Inconsistent formats:} 
The users may employ different formats to compose bug reports depending on their writing preferences.
Common formats include numeric indexing, list indexing, or using commas in a line.
The baselines can correctly separate the steps following these patterns.
However, regarding the users' understanding of the bugs, they may compose the steps under the same state such as ``->'' and ``then'' as shown in Figure~\ref{fig:rq1}-A, but the baseline approaches fail to interpret them as separate steps.

\renewcommand{\arraystretch}{1.05}
\begin{table}
    \footnotesize
    \tabcolsep=0.12cm
	\centering
	\caption{Performance comparison of S2R extraction.}
	\label{tab:rq1_performance}
	\begin{tabular}{l||c|c|c|c|c} 
	    \hline
	    \multirow{2}{*}{\bf{Method}} & \multirow{2}{*}{\makecell{\bf{Acc. of} \\ \bf{Step Extraction}}} & \multicolumn{4}{c}{\bf{Acc. of Entity Extraction}} \\
	    \cline{3-6}
	     & & Action & Component & Input & Direction \\
	     \hline
            ReCDroid & 49.5\% & 65.9\% & 45.4\% & 30.8\% & 42.9\% \\
            MaCa & 51.1\% & 70.1\% & 41.6\% & 33.3\% & 50.0\% \\
		\hline
            \tool w/o Few & 83.9\% & 81.9\% & 84.5\% & 87.1\% & 85.7\% \\
            \tool w/o CoT & 86.9\% & 83.6\% & 87.6\% & 87.1\% & 92.9\% \\
            \hline
            \tool & \bf{90.4\%} & \bf{91.1\%} & \bf{90.0\%} & \bf{89.7\%} & \bf{92.9\%} \\
            \hline
	\end{tabular}
\end{table}

\textit{2) Dependent context.}
Context dependency can occur from a macro perspective in the bug reports.
For example, in Figure~\ref{fig:rq1}-B, the users demonstrate an action of ``deleting'' a value, which is equivalent to ``leave the EditText empty'', thus, they should be interpreted as one step.
In Figure~\ref{fig:rq1}-C, the users may omit duplicate steps by describing them as ``repeat the previous step'' or ``create another one'', etc.
However, without syntactic awareness, the baseline approaches do not monitor the relationships among the words in a sentence, causing inaccurate step extraction.

Context dependency can also occur at a more granular level.
The users may use conjunction words to order the reproduction steps that differ from their syntactic order.
The baseline MaCa proposes a heuristic approach based on constituency grammar to deal with common connective words (i.e., ``before'', ``after'', ``when'', etc.), but it is limited to less commonly-used words.
For instance, in Figure~\ref{fig:rq1}-D, the step ``Favorites settings'' should happen first, and then the step ``Show favorites above search bar'', due to the semantic of the connective word ``under'' between the steps, which the baselines fail to recognize.

\textit{3) Diversified words:}
The users may use different words or phrases to describe the same concept, making it difficult for baseline approaches to recognize semantic patterns.
This can include synonyms, alternative phrasing, or even colloquial language.
For example, as shown in Figure~\ref{fig:rq1}-E, the users colloquially use ``go to'', ``click'', ``choose'', ``open'', and ``select'' to refer to the action of ``tap''.
Even for the same word, the user's choice of expression can affect the recognition, e.g., ``click the button'' and ``the button is clicked'' in Figure~\ref{fig:rq1}-F.
While the baselines have identified some common vocabulary of words, the inconsistency of human language still poses challenges in recognizing diversified word patterns in bug reports.

Albeit the good performance of our approach, we still make the wrong extraction for some bug reports. 
We manually check those incorrect S2R entity extractions and identify one common cause.
The users may write the S2Rs ambiguously, causing the LLMs to fail to capture the user intent.
For example, in Figure~\ref{fig:rq1}-G, the users aim to set the key \textbf{to} ``shop'' (i.e., [Input] [``key''] [``shop'']) and value to ``*'' (i.e., [Input] [``value''] [``*'']).
However, due to the missing \textbf{to} keyword, LLMs misunderstand the S2R and extract the step as [Input] [``shop''] [``*''].
Note that the LLMs can correctly extract the S2Rs by adding the \textbf{to} keyword, suggesting that our performance could be further improved with well-written bug reports.


%% file: eval_rq2.tex
\subsection{RQ2: Performance of Guided Replay}
\label{sec:rq2}
\textbf{Experimental Setup.}
To answer RQ2, we evaluated the ability of our approach (in Section~\ref{sec:phase2}) to accurately map the entities in the S2Rs to the components in the GUIs for bug reproduction.
We used our experimental dataset (outlined in Section~\ref{sec:rq1}), containing 88 distinct textual S2Rs.
However, many S2Rs are no longer reproducible due to two reasons.
First, many real-world bugs have been fixed and the apps have been patched, making it difficult to find the corresponding previous versions of the apps for reproduction. 
Second, some bugs (e.g., financial, social apps) require extensive information like authentication, database, or specific hardware, which are beyond the scope of this study.
To that end, we asked the two students to locate the corresponding apps and interact with them to ensure the S2Rs were reproducible.
If the students could not reproduce a bug report, the first author would further assess its reproducibility. 
As a result, we collected 48 reproducible S2Rs from the bug reports.

\textbf{Metrics.}
To measure the performance of our approach, we employed reproducibility as the evaluation metric, i.e., whether the method can successfully reproduce the bugs. 
The higher the reproducibility score, the better the approach can replicate the S2Rs on the GUIs to trigger the bugs.

\textbf{Baselines.}
We adopted the state-of-the-art method \textit{ReCDroid}~\cite{zhao2019recdroid} as the baseline to compare with our method.
Specifically, ReCDroid employs a greedy algorithm to search for the matched components on the GUIs by using the word embedding similarity metric Word2Vec.
In addition, we adopted two derivations of our approach as the ablation baselines to demonstrate the advantages of \tool, including without few-shot learning (\textit{\tool w/o Few}), and without chain-of-thought (\textit{\tool w/o CoT}).
Note that we did not adopt MaCa~\cite{liu2020automated} as the baseline, as it solely focuses on S2R entity extraction without including bug replay.

\renewcommand{\arraystretch}{1.05}
\begin{table}
    \scriptsize
    \tabcolsep=0.11cm
	\centering
	\caption{Performance comparison of guided replay.}
	\label{tab:rq2_performance}
	\begin{tabular}{l|r||c|c|c|c} 
	    \hline
	    \multirow{2}{*}{\bf{Bug}} & \multirow{2}{*}{\bf{No. Steps}} & \multicolumn{4}{c}{\bf{Method}} \\
	    \cline{3-6}
	     & & ReCDroid & \tool w/o Few & \tool w/o CoT & \textbf{\tool} \\
	     \hline
           \rowcolor{lightgray} Anki\#4586 & 3 & \cmark & \cmark & \cmark & \cmark\\
            Anki\#5638 & 4 & \xmark & \xmark & \cmark & \cmark \\
            \rowcolor{lightgray} Anki\#2564 & 3 & \xmark & \cmark & \cmark & \cmark \\
            Anki\#3224 & 2 & \xmark & \xmark & \xmark & \cmark \\
            \rowcolor{lightgray} Anki\#10584 & 1 & \cmark & \cmark & \cmark & \cmark \\
            AnyMemo\#440 & 5 & \cmark & \cmark & \cmark & \cmark \\
            \rowcolor{lightgray} Birthdroid\#13 & 1 & \cmark & \xmark & \xmark & \cmark \\
            NewsBlur\#1053 & 5 & \cmark & \cmark & \cmark & \cmark \\
            \rowcolor{lightgray} LibreNews\#22 & 5 & \cmark & \cmark & \cmark & \cmark \\
            LibreNews\#23 & 7 & \cmark & \cmark & \cmark & \cmark \\
            \rowcolor{lightgray} LibreNews\#27 & 4 & \xmark & \xmark & \xmark & \xmark \\
            Transistor\#63 & 5 & \cmark & \cmark & \cmark & \cmark \\
            \rowcolor{lightgray} ScreenCam\#25 & 4 & \cmark & \cmark & \cmark & \cmark \\
            News\#487 & 2 & \cmark & \xmark & \xmark & \cmark \\
            \rowcolor{lightgray} k9mail\#3255 & 4 & \cmark & \cmark & \cmark & \cmark \\
            k9mail\#2612 & 3 & \cmark & \xmark & \xmark & \xmark \\
            \rowcolor{lightgray} k9mail\#2019 & 1 & \cmark & \cmark & \cmark & \cmark \\
            openMF\#734 & 2 & \cmark & \cmark & \cmark & \cmark \\
            \rowcolor{lightgray} FamilyFinance\#1 & 6 & \xmark & \xmark & \xmark & \cmark \\
            trickytripper\#42 & 4 & \xmark & \xmark & \xmark & \cmark \\
            \rowcolor{lightgray} NoadPlayer\#1 & 4 & \cmark & \cmark & \cmark & \cmark \\
            calendula\#134 & 2 & \cmark & \cmark & \cmark & \cmark \\
            \rowcolor{lightgray} StreetComplete\#1093 & 3 & \xmark & \xmark & \xmark & \xmark \\
            OmniNotes\#592 & 3 & \cmark & \cmark & \cmark & \cmark \\
            \rowcolor{lightgray} Markor\#1020 & 2 & \xmark & \cmark & \cmark & \cmark \\
            Markor\#331 & 6 & \xmark & \xmark & \xmark & \xmark \\
            \rowcolor{lightgray} KISS\#1481 & 5 & \xmark & \cmark & \cmark & \cmark \\
            ultrasonic\#187 & 3 & \xmark & \cmark & \cmark & \cmark \\
            \rowcolor{lightgray} andOTP\#500 & 3 & \xmark & \xmark & \xmark & \xmark \\
            k9mail\#3971 & 3 & \xmark & \cmark & \cmark & \cmark \\
            \rowcolor{lightgray} kiwix\#1414 & 4 & \cmark & \cmark & \cmark & \cmark \\
            kiwix\#555 & 2 & \cmark & \cmark & \cmark & \cmark \\
            \rowcolor{lightgray} qksms\#1155 & 3 & \xmark & \xmark & \xmark & \xmark \\
            Aegis\#287 & 3 & \xmark & \xmark & \xmark & \cmark \\
            \rowcolor{lightgray} AmazeManager\#1796 & 5 & \xmark & \xmark & \xmark & \cmark \\
            ActivityDiary\#285 & 6 & \xmark & \xmark & \cmark & \cmark \\
            \rowcolor{lightgray} APhotoManager\#116 & 3 & \xmark & \xmark & \xmark & \xmark \\
            collect\#3222 & 2 & \xmark & \cmark & \cmark & \cmark \\
            \rowcolor{lightgray} commons\#2123 & 3 & \xmark & \xmark & \xmark & \xmark \\
            FirefoxLite\#5085 & 4 & \xmark & \cmark & \cmark & \cmark \\
            \rowcolor{lightgray} nextcloud\#1918 & 5 & \cmark & \cmark & \cmark & \cmark \\
            nextcloud\#5173 & 4 & \xmark & \cmark & \cmark & \cmark \\
            \rowcolor{lightgray} osmeditor\#729 & 2 & \cmark & \cmark & \cmark & \cmark \\
            osmeditor\#637 & 4 & \xmark & \xmark & \xmark & \xmark \\
            \rowcolor{lightgray} WordPress\#11135 & 3 & \xmark & \cmark & \cmark & \cmark \\
            WordPress\#10302 & 3 & \cmark & \cmark & \cmark & \cmark \\
            \rowcolor{lightgray} CineLog\#60 & 5 & \xmark & \xmark & \xmark & \cmark \\
            AndrOBD\#144 & 3 & \xmark & \xmark & \xmark & \cmark \\
            \hline
            \hline
            \rowcolor{darkgray} Reproducibility & - & 45.8\% & 58.3\% & 62.5\% & \textbf{81.3\%} \\
            \hline
	\end{tabular}
\end{table}

\begin{figure}
	\centering
	\includegraphics[width = 0.99\linewidth]{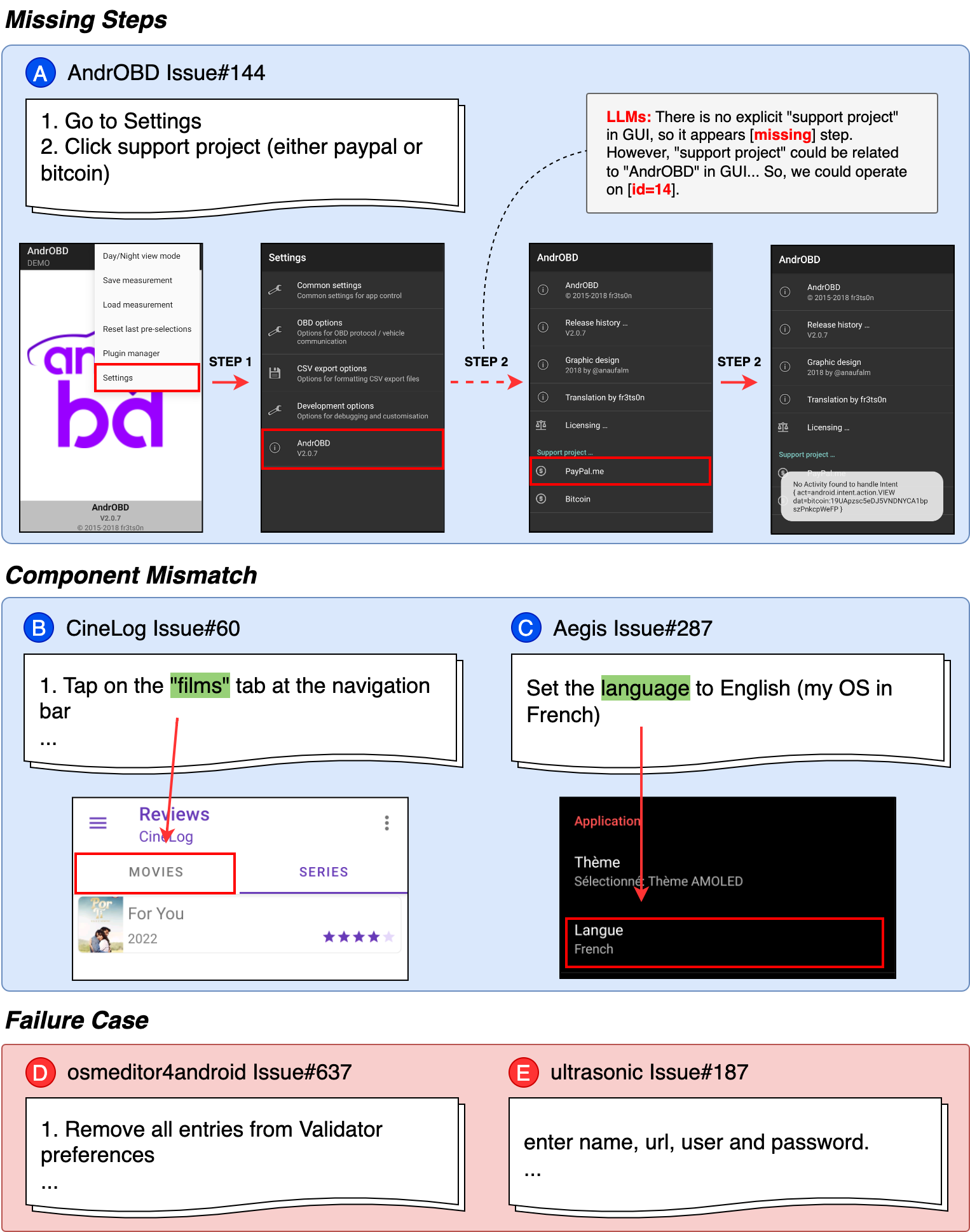}
	\caption{Examples of guided replay.}
	\label{fig:rq2}
\end{figure}

\textbf{Results.}
Table~\ref{tab:rq2_performance} shows detailed results of the reproducibility for each bug.
Our approach outperforms all the baselines and ablations, i.e., on average 81.3\% compared with 45.8\%, 58.3\%, and 62.5\%, for ReCDroid, \tool w/o Few, and \tool w/o CoT, respectively.
We observe that chain-of-thought leads to a substantial improvement (i.e., 18.8\% boost) in the performance of \tool, indicating that the LLMs can better understand the task by processing it step-by-step.

We further conduct a qualitative analysis to compare the capabilities of our approach with the baselines.
Overall, we summarize two common reasons which can be seen in Figure~\ref{fig:rq2}.

\textit{1) Missing steps:}
It is a key factor in the failure of automated bug reproduction in baselines.
This issue may arise when users overlook some "unimportant" steps.
Without a complete set of instructions, automated approaches fail to reproduce the steps needed to trigger the bugs.
In contrast, our approach exploits the semantic understanding of subsequent steps and the components in the current GUI to potentially identify the most probable actions that can restore the missing steps.
For instance, in Figure~\ref{fig:rq2}-A, since there is no explicit target component (``support project'') in the current GUI, the LLMs can identify the potential component (``AndrOBD'', which is the project name) due to their semantic correlation.

\textit{2) Component mismatch:}
Another common reason that makes it difficult for baselines to repeat the step is the component mismatch.
There may be two possible reasons why an S2R cannot directly match components on the GUI.
First, the users may describe the component imprecisely.
For instance, in Figure~\ref{fig:rq2}-A, the ``movies'' button is described as ``films'', making it challenging for automated approaches to identify the exact GUI component the S2R refers to.
Second, for multi-lingual apps, the languages of S2Rs and GUIs may be different.
Figure~\ref{fig:rq2}-B shows an example that the user reports the bug in English while it is reproduced on a GUI in French, causing the reproduction to fail.
In contrast, our approach leverages LLMs, which capture a large-scale corpus and language resources, to understand the morphological forms of components including abbreviations, synonyms, misspellings, and languages, and search for the most relevant component in the GUI.

Although our approach demonstrates good bug reproducibility, we still fail to reproduce some unclear S2Rs.
For example, the users may not describe the specific values for input fields, as seen in Figure~\ref{fig:rq2}-D.
To address it, we generate random text for the input fields, i.e., the ``test'' value for the name field.
However, some input fields require valid values, i.e., the URL field requires an input value starting with ``https://''.
This could be achieved by understanding the context information of the input fields and designing linguistic patterns to prompt LLMs to generate valid input values.
We would like to put it as our future work.


%% file: eval_rq3.tex
\subsection{RQ3: Efficiency Performance}
\textbf{Experimental Setup.}
To answer RQ3, we evaluated the overhead of our approach by calculating the average time it takes for a bug report to pass through each of the two phases of the \tool approach.
All of our experiments were conducted on the commodity hardware, a 2.6 GHz Macbook Pro with 6 dedicated CPU Intel Core as the \tool engine, and the official Android x86-64 emulator as the device server.
Our approach could perform substantially faster on specialized hardware.
Note that, inferring the entities and guidance from ChatGPT (our underlying LLM implementation) can be unstable and highly dependent on the ChatGPT server and user volume.
Therefore, we ran the inference three times for our approach and the ablation baselines, using the average time for performance evaluation.

\textbf{Metrics.}
To measure the overhead of \tool, we employed the time it takes for both extracting S2R entities from the bug report and guiding the bug replay.
The less time it takes, the more efficiently the method can reproduce the bugs.

\textbf{Baselines.}
We used \textit{ReCDroid} and \textit{MaCa} as the baselines, as well as \tool without few-shot and \tool without chain-of-thought as the ablation studies of our approach.

\textbf{Results.}
Table~\ref{tab:rq3_performance} shows the performance comparison with the baselines.
Our approach takes \tool 255.75 seconds on average to reproduce a bug report, i.e., 2.11 seconds to extract the S2R entities from the bug report and 253.64 seconds to replay the S2R in the device.
In comparison, it takes the ReCDroid method on average 1360.92 seconds, indicating the efficiency of our approach which saves nearly 5x time per bug replay.
Notably, our approach accelerates 435\% time in guiding replay in the GUI screen compared with ReCDroid.
This is primarily due to missing steps in the S2Rs, which cause ReCDroid to randomly explore the app with repeated back-and-forth exploration, which is a time-consuming process.
In contrast, our approach elicits the LLMs to understand the semantics of the S2R context and current GUI to predict the most probable components for bridging the missing steps.
We expect that the overhead of our approach could be further accelerated by more advanced hardware.

\renewcommand{\arraystretch}{1.05}
\begin{table}
    \footnotesize
    \tabcolsep=0.25cm
	\centering
	\caption{Performance comparison of average runtime per bug report.}
	\label{tab:rq3_performance}
	\begin{tabular}{l|c|c} 
	    \hline
	    \bf{Method} & \bf{S2R Entity Extraction (sec)} & \bf{Guided Replay (sec)} \\
	     \hline
            ReCDroid & 3.75 & 1357.17 \\
            MaCa & 2.18 & - \\
            \tool & \bf{2.11} &  \bf{253.64} \\
            \hline
	\end{tabular}
    \vspace{-0.3cm}
\end{table}


%% file: userstudy.tex
\subsection{RQ4: Usefulness of \tool}
\label{sec:userstudy}
\textbf{Experimental Setup.}
To investigate the perceived usefulness of \tool, we conducted a small-scale user study with eight participants, including five graduate students and three app developers.
The participants were recruited from the university’s internal slack channel or through direct contact with the authors. 
Given that the graduate students all have at least 1.5 years of experience in Android app development and bug replay.
The three app developers are more professional and have more than two-year working experience in Android development in the industry.

\textbf{Procedure:}
To mitigate the threat of user distraction, we conducted the experiment in a quiet room individually without mutual discussion.
We interviewed the participants using a set of questions organized in two sessions. 
The first session aimed to collect information on participants’ backgrounds, including their role at the company, the frequency of addressing bug reports and replaying bugs, the challenges they face in their practices, etc. 
The second session aimed to assess \tool’s potential usefulness in replaying the S2Rs from bug reports. 
Each participant was asked to reproduce the same set of 5 randomly selected bug reports from issue trackers and GitHub.
These bug reports were of diverse difficulty ranging from 2 to 8 S2Rs in the bug reports.
To minimize the impact of stress, participants were allowed to take a short break between each bug replay.
We only recorded the time used to reproduce the bug reports, including understanding the textual S2Rs and replicating the steps in Android.
Note that all the experiment apps were pre-installed.
Participants had up to 10 minutes for each bug replay.
At the end of the tasks, we showcased the bug replay automatically generated by our tool \tool and asked participants for feedback, including a 5-Likert scale question about their preference for using our tool, suggestions for improvement, etc.

\textbf{Results.}
Table~\ref{tab:user_study} shows the detailed experiment results.
Although most participants can successfully finish the bug replay on time, \tool reproduces the bug report much faster than that of participants (with an average of 480.7 seconds versus 269.4 seconds). 
In fact, the average time for participants' reproduction is underestimated, because they fail to reproduce 3 bugs within 10 minutes, which means that participants may need more time for bug replay. 
In contrast, our automated approach finishes all the tasks within 7 minutes.

\renewcommand{\arraystretch}{1.05}
\begin{table}
    \footnotesize
    \tabcolsep=0.125cm
	\centering
	\caption{Performance comparison of usefulness evaluation.}
	\label{tab:user_study}
	\begin{tabular}{l|r||c|c|c|c} 
	    \hline
	    \multirow{2}{*}{\bf{Bug}} & \multirow{2}{*}{\bf{No. Steps}} & \multicolumn{2}{c|}{\bf{Participants}} & \multicolumn{2}{c}{\bf{\tool}} \\
	    \cline{3-6}
	     & & Success & Time (sec) & Success & Time (sec) \\
	     \hline
            KISS & 2 & 75\% & 423.1 & 100\% & 262.9 \\
            NeuroLab & 2 & 100\% & 396.5 & 100\% & 228.0 \\
            Alibaba & 5 & 87.5\% & 545.7 & 100\% & 301.2\\
            Wechat & 7 & 100\% & 493.3 & 100\% & 205.1 \\
            Anki & 8 & 100\% & 544.8 & 100\% & 349.7 \\
            \hline
            \hline
            Average & - & 92.5\% & 480.7 & \bf{100\%} & \bf{269.4} \\
            \hline
	\end{tabular}
\end{table}

After observing the bug replay automatically generated by \tool, all participants strongly favored using \tool in practice, with an average preference score of 4.5 out of 5.0.
To gain insight into the usefulness of our approach, we collect feedback from the participants and summarise two practical challenges of manual replay.
First, understanding the S2Rs from the bug reports is surprisingly time-consuming, as it involves grasping the context, reordering the steps, analyzing the potential actions and components, etc.
Second, it is difficult to determine the trigger for the missing steps, resulting in participants' guesses of the action for triggering the next steps.
That trial-and-error makes the bug replay process tedious and time-consuming.
It is especially severe for junior developers who are not familiar with the app.

Participants also provide valuable feedback to help improve \tool.
For example, one participant suggests a strategy for interaction between developers and LLMs for bug replay by adding a confidence value to the LLMs.
If the inference is unreliable (i.e., the confidence value is low), there will be an interactive conversation to ask developers to confirm the operations.
This strategy could efficiently and effectively improve the performance of bug reproduction with human-AI collaboration.
We will investigate the possible solution in our future work.

%% file: threats.tex
\section{THREATS TO VALIDITY}
In our experiments evaluating our model, threats to internal validity may arise from the randomness of LLMs generation, which may generate different results for different runs.
Namely, across different runs of the same prompt, the obtained metrics could vary.
To mitigate this threat, we ran the LLM-related approaches (i.e., \tool and ablation baselines) three times and the metrics were then from the aggregation of the three runs.
Another potential confounding factor concerns the manual labeling of the ground-truth for S2R entity extraction (RQ1).
To mitigate any potential subjectivity or errors, we gave the annotators a training session and a passing test before labeling.
We asked them to independently annotate without any discussion and came to a consensus on the finalized ground-truth.

The main external threat to the validity of our work is the representative of the testing dataset selected to evaluate our approach.
To mitigate this threat, we collected bug reports from previous related research, representing an unbiased testing dataset for our study.
In addition, we randomly collected five practical bug reports to further evaluate the usefulness of our tool with five graduate students and three developers in real-world development environments.
A potential confounding factor concerns the representation of graduate students in the user study.
To mitigate this threat, five graduate students have at least 1.5-year experience in Android app development and bug reproduction, so they are recognized as substitutes for developers in software engineering research experiments~\cite{salman2015students}.

%% file: related.tex
\section{Related Work}
We review the related work in two main areas: 1) bug record and replay, and 2) large language models for software engineering.

\subsection{Bug Record and Replay}
Nurmuradov et al.~\cite{nurmuradov2017caret} introduced a record and replay tool for Android applications that captures user interactions by displaying the device screen in a web browser. 
It used event data captured during the recording process to generate a heatmap that allows developers to reproduce how users are interacting with an application. 
Additional tools including ECHO~\cite{sui2019event}, Reran~\cite{gomez2013reran}, Barista~\cite{ko2006barista}, and Android Bot Maker~\cite{web:botmaker} are program-analysis-related applications for the developers to record and replay user interactions. 
However, they required the installation of underlying frameworks such as replaykit~\cite{web:replaykit}, troyd~\cite{jeon2012troyd}, or instrumenting apps which are too heavy for end users.
In contrast to these approaches, our \tool is rather lightweight which just requires natural language descriptions of the bug reproduction steps.

Many previous research works have focused on processing bug reproduction steps to automate bug replay.
For example, Fazzini et al.~\cite{fazzini2018automatically} proposed Yakusu, a program analysis and natural language processing tool to generate executable test cases from bug reports.
Zhao et al.~\cite{zhao2019recdroid} further improved the tool namely ReCDroid, by leveraging the lexical knowledge in the bug reports to automatically reproduce crashes.
However, these works did not consider the temporal order of S2Rs (e.g., when, after, etc.).
Liu et al.~\cite{liu2020automated} proposed MaCa, a natural language processing (NLP) and machine learning (ML) approach to normalize the S2Rs in the bug reports and extract the S2R entities more effectively.
However, these approaches highly depended on the S2R writing in the bug report including formatting, word usage, and granularity~\cite{erfani2014works,bettenburg2008makes,koru2004defect,cao2014symcrash}, which constrains its deployment in the real world. 
In contrast, our approach does not construct any sophisticated manually-crafted patterns and pre-defined vocabulary of words.
We leverage a single model LLMs with novel prompt engineering to elicit the natural language understanding and logical reasoning to extract S2R entities and reproduce the bugs.


Some works attempted to assist reporting process~\cite{feng2021auto,feng2022auto,xie2020uied,xie2022psychologically,chen2023unveiling,feng2023video2action} and improve the bug report quality~\cite{moran2015auto,wong2014boosting,feng2022efficiency,feng2023read}.
For example, Chaparro et al.~\cite{chaparro2017detecting} developed DeMIBuD to detect missing information from bug reports.
Fazzini et al.~\cite{fazzini2022enhancing} proposed a mobile application EBug, suggesting accurate reproduction steps while the users are writing them. 
More recently, Yang et al.~\cite{song2022toward} proposed an interactive chatbot system BURT to guide users to report essential bug report elements.
These approaches are complementary to our approach, as they increase bug report quality at the time of reporting which in turn, could potentially improve the capabilities of LLMs' understanding.

\subsection{Large Language Models for Software Engineering}
Following the success of LLMs in many natural language processing tasks, researchers have started exploring their potential in software engineering. 
Several studies have focused on automated code completion~\cite{xu2022systematic,huang2022prompt,li2022competition,jain2022jigsaw,deng2022fuzzing}. 
More recently, Codex~\cite{web:codex} employed by Github Copilot has shown promising results in generating code from natural language comments, which has the potential to significantly reduce developers' coding efforts, although some minor issues still exist.
This has sparked interest in using LLMs to resolve code issues through program patching~\cite{kolak2022patch,fan2022improving,jiang2021cure,fu2022vulrepair}. 
For instance, AlphaRepair, introduced by Xia et al.~\cite{xia2022less}, effectively encoded the intricate relations between patches and the contexts of codes using LLMs. 
Additionally, some studies have examined the impact of LLMs on software development~\cite{liu2022fill,imai2022github}, developer productivity~\cite{vaithilingam2022expectation,ziegler2022productivity,dakhel2022github}, and security vulnerabilities~\cite{pearce2021empirical,grishina2022enabling,pearce2022examining}. 
LLMs remain an ongoing research topic in the software engineering community. Our work opens up a new possibility for integrating LLMs and GUI understanding, enabling LLMs to understand S2Rs and guide dynamic GUI screens to automated bug reproduction.

%% file: conclusion.tex
\section{Conclusion}
Inspired by the success of Large Language Models (LLMs) in natural language understanding, this paper proposes \tool, a lightweight approach to automatically reproduce the bugs from the S2Rs based on prompt engineering.
In detail, we first prompt the LLMs by giving entity specifications, a few representative exemplars, and developers' step-by-step reasoning, to help LLMs elicit the knowledge to extract the entities from the S2Rs in the bug report like a developer expert.
Given each step entity, we prompt the LLMs by giving the current GUI screen and few-shot learning with chain-of-thought reasoning to dynamically guide the probable target components to reproduce the bugs.
The experiments and user study demonstrate the effectiveness, efficiency, and usefulness of our approach in accelerating bug reproduction.

In the future, we will keep improving our \tool for better effectiveness.
For example, a bug report usually contains a wealth of information, including stack trace, error logs, screenshots, screen recordings, etc.
We could take this information into the consideration to enhance the understanding of LLMs to the bugs.